\documentstyle[epsf,rotate]{aa}

\begin{document}
\newcommand{\HII}{\mbox{H\,{\sc ii}}}
\newcommand{\OII}{\mbox{O\,{\sc ii}}}
\newcommand{\LCO}{\mbox{L$'_{\sc CO}$}}

\thesaurus{11.06.1, 11.09.4 11.19.3}
\title{CO detection of the extremely red galaxy HR10}

\offprints{P. Andreani}

\author{
   Paola Andreani\inst{1}
\and
   Andrea Cimatti\inst{2}
\and
   Laurent Loinard\inst{3}
\and
Huub R\"ottgering\inst{4}
}
   
\institute{Osservatorio Astronomico di Padova
     vicolo dell'Osservatorio 5, I-35122 Padova, Italy,
     e-mail: andreani@pd.astro.it\\
     Present-address: Max-Planck I. f. extraterrestrische
     Physik, Postfach 1603, D-85740 Garching, Germany
\and
     Osservatorio Astrofisico di Arcetri, 
     Largo E. Fermi 5, I-50125 Firenze, Italy,
     e-mail: cimatti@arcetri.astro.it
\and
     Institut de Radioastronomie Millim\'etrique,
     300, rue de la Piscine, St. Martin d'H\`eres, France
     e-mail: loinard@iram.fr
\and
     Sterrewacht Leiden, Sterrewacht, Postbus 9513, Leiden 2300 RA
     the Netherlands e-mail: rottgeri@strw.leidenuniv.nl
}

\date{Received ...; accepted ....}

\titlerunning{CO detection of HR10}
\authorrunning{Andreani et al.}
\maketitle

\begin{abstract}
CO $J = 5 - 4$ and $J = 2 - 1$ emission lines were detected 
towards the extremely red galaxy (ERG) HR10 (J164502+4626.4) at $z=1.44$.
The CO intensities
imply a molecular gas mass M(H$_{\rm 2}$) of $1.6 \times 10^{11}$ h$^{-2}_{\rm 50}$ 
M$_\odot$, and, combined with the intensity of the dust continuum,
a gas-to-dust mass ratio around 200-400 (assuming galactic values for the
conversion factors). The peak of the CO lines are 
at the same redshift as the [O{\scriptsize II}]3727 line, but blue-shifted by
430 km s$^{-1}$ from the H$\alpha$ line. These CO detections confirm the 
previous results that HR10 is a highly obscured object with a large thermal 
far--infrared luminosity and a high star--formation rate. The overall 
properties of HR10 (CO detection, L$_{\rm FIR}$ to L$^\prime_{\rm CO}$ ratio,
and FIR to radio flux ratio) clearly favour the hypothesis 
that its extreme characteristics are related to
star--formation processes rather than to a hidden AGN.

\keywords{ISM: molecules - Galaxies: formation - Galaxies: individual (HR10)}
\end{abstract}

\section{Introduction}

The recent detections of CO emission at cosmological distances provide hints 
about the physical structure of newly formed objects (\cite{com99} and 
reference therein). A measure of the total gas and dust mass is indeed a 
very useful indicator of the object evolutionary status, because it provides 
an estimation of the fraction of the galaxy which has yet to be turned into 
stars at the epoch of observation.
At high redshifts such measurements, therefore, provide hints about
the occurrence of active star-formation processes and help
in investigating models of galaxy formation (\cite{sil97}).

Only a handful of distant objects was detected so far in CO, and most of them 
appear to be magnified by gravitational lenses (table 1). For these objects
the mass of molecular gas inferred from the CO intensities -- corrected for 
gravitational amplification and using the Galactic CO to H$_{\rm 2}$ conversion 
factor -- turns out to be $30 \div 80$ \% of the total dynamical mass, with 
typical values of M(H$_{\rm 2}$) $\simeq 10^{10} \div 10^{11}$ M$_\odot$. These 
masses could be somewhat smaller if the CO to H$_{\rm 2}$ conversion factor were 
higher than in the Galaxy.
Large quantities of CO are expected when large FIR luminosities
and dust content are detected. Most of the 
objects detected until now in CO were indeed selected because they were 
dust--rich systems.

\begin{table*}
\caption{List of high--redshift objects detected to date in CO.}
\begin{center}
\begin{tabular}{||l|l|c|c|l||}
\hline
& & & & \\
 \multicolumn{1}{||c|}{Name} & \multicolumn{1}{|c|}{type}
& \multicolumn{1}{|c|}{redshift}
& \multicolumn{1}{|c|}{M$_{\rm H_2}$ ($10^{11}$ M$_\odot$)}
& \multicolumn{1}{|c||}{references} \\
& & & & \\
\hline
& & & & \\
F10214+4724 $^l$ & ULIRG & 2.28 & 4.2$^\star$ &\cite{sol92} \\
Cloverleaf $^l$ & QSO & 2.558 & 0.9$^\star$ &\cite{bar94} \\
BR1202-0725 $^l$ & QSO & 4.69 & 2.4 &\cite{oht96,omo96} \\
MG0414+0534 $^l$ & QSO & 2.639 & 3.3 &\cite{bar98} \\
SMM02399-0136 $^l$ & submm-HLG & 2.808 & 1.8$^\star$& \cite{fra98} \\
SMM14011+0252 $^l$ & submm-HLG & 2.565 & 1.1$^\star$ &\cite{fra99} \\
APM08279+5255 $^l$ & BAL quasar & 3.911 & 0.03$^\star$ & \cite{dow99}\\
BRI1335-0417 & QSO & 4.407 & 2.4&\cite{gui97} \\
53W002 & weak RG & 2.39 & 1.3 & \cite{sco97a,all99} \\
4C 60.07 & RG & 3.79 & 1.8 & \cite{pap99} \\
6C 1909+722  & RG & 3.53 & 1.0 &\cite{pap99}\\
\hline
\end{tabular}
\end{center}
$l$: evidence for a lensed object\par
$\star$: corrected for magnification
\end{table*}

HR10 belongs to the class
of objects with very red colours ($R-K>6$). Their faintness
at optical/NIR wavelengths makes the redshift determination and
the investigation of their nature difficult
even with 4m class telescopes.
HR10 is so far the only one with a measured spectroscopic redshift
(\cite{gra96}). One of the main issues regarding ERGs is whether they are
young and starbursting galaxies hidden in the optical by a large amount of dust
or whether they are old passively evolving galaxies
at $z\geq 1$. Recent results indicate that both classes contribute to
the population of ERGs (\cite{cim99b}).\hfill\break
HR10 was first detected in the submm/mm continuum
with the IRAM 30m equipped with the MPIfR bolometer and 
with the JCMT equipped with the SCUBA double arrays (\cite{cim98}). Subsequent 
observations confirmed the submm detection of this galaxy (\cite{dey99}). The 
inferred properties of this object show its extraordinary nature: its dust 
mass is $4 \div 8 \times 10 ^8$ M$_\odot$ and its total FIR 
luminosity in the range 10 --2000 $\mu$m rest--frame is $2 \div 2.5 \times 10^{12}$ 
L$ _\odot$ (H$_{\rm 0} =50$ Mpc/km/s). This places HR10 in the class of
the ultra-luminous infrared 
galaxies and suggests the presence of a star--forming object with a SFR of 
$\sim 200 \div 500$ M$_\odot$ yr$^{-1}$ or even higher.

In this paper we present CO(2-1) and CO(5-4) observations of HR10 made with 
the IRAM Plateau de Bure interferometer. The observations are described in 
Sect. 2, the resulting detection of both lines are presented in Sect. 3, 
while implications of these measurements are reported in Sect. 4. Throughout 
the paper, we adopt H$_{\rm 0} = 50$ Mpc/km/s and q$_{\rm 0} = 0.5$.

\section{Observations}

The observations were done partly during the winter 1998-99 and partly during
the summer of 1999 with the IRAM Plateau de Bure Interferometer. A dual SIS
3mm/1.3mm receiver was used to observe simultaneously the CO(2-1) (redshifted
in the 3mm band) and the CO(5-4) (redshifted in the 1.3mm band) lines. In 
both cases, the 3mm receiver was connected to two units of the Correlator, 
providing a velocity coverage of about 850 km s$^{-1}$, while 4 units were 
connected to the 1.3mm receiver providing a velocity coverage of about 650 
km s$^{-1}$.
\hfill\break
The observations made during the winter 1998-99 assumed a redshift of $z= 
1.443$ deduced from the H$\alpha$ line. This yields frequencies of $\nu =
94.405$ GHz for the CO(2-1) line and $\nu = 235.982$ GHz for the CO(5-4) line. 
Both lines were detected, but blue-shifted by about 400 km 
s$^{-1}$ with respect to the H$\alpha$ line and thus truncated.
The observations were repeated during the summer 1999 with a new 
frequency setup: $\nu = 94.531$ GHz and $\nu = 236.297$ GHz for the CO(2-1) 
and CO(5-4) lines respectively.
\hfill\break
The observations were made in the CD configuration of the interferometer.
The phase drifts were calibrated by observing the nearby quasars 1633+382
and 1732+389 every 20 minutes throughout the observations. The amplitude was 
calibrated with the quasars 3C273, 3C345 and 2145+067, and the compact 
\HII\ region MWC349 at the beginning and/or the end of each transit. The
passband of the system was calibrated using 3C273 or 2145+067.
At 1.3mm, the total useful integration time on source was about 9 hours 
during the winter 1998/99, and 12 hours during the summer 1999. More 
integration time was available at 3mm because some additional data were
obtained, marginal at 1.3mm but of good quality at 3mm.
\hfill\break
The 3mm receiver was tuned in single side band, while the 1.3mm was tuned 
in double side band, with the CO(5-4) in the lower side band. The upper
side band was then available to measure the continuum.

Three visibility tables were produced from the calibrated data: two spectral
tables containing the CO(2-1) and CO(5-4) data respectively and one continuum
table with the 1.3mm continuum.
The spectral data were smoothed in frequency to improve their
signal-to-noise ratio: the velocity resolution of the CO(2-1) data was 
smoothed to 50 km s$^{-1}$ and that of the CO(5-4) to 75 km s$^{-1}$. The
UV tables were imaged, CLEANed and restored with elliptical gaussian beams
of $5.6^{\prime \prime} \times 4.6^{\prime \prime} $ ($PA = 56^{\circ}$)
and $2.9 ^{\prime \prime} \times 2.0^{\prime \prime} $
($PA = 65^{\circ}$) for the 3mm and 1.3mm data respectively.

\begin{figure*}
\epsfysize=15cm
{\epsffile{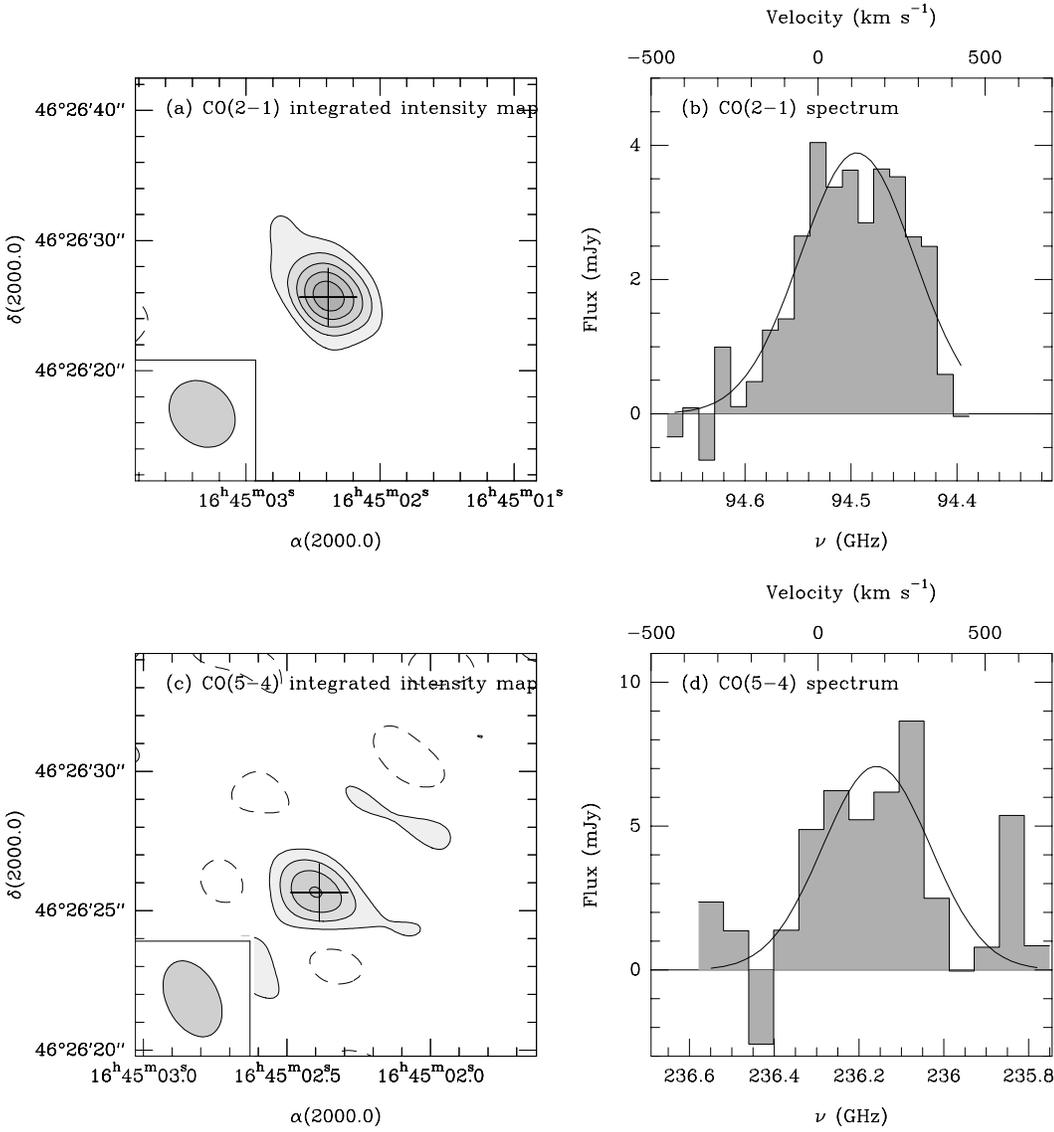}}
\caption[]{{\bf (a)} CO(2--1) integrated intensity map; the contours are at 
0.25 Jy beam$^{-1}$ km s$^{-1}$. {\bf (b)} CO(2--1) spectrum at the center 
of the source. {\bf (c)} CO(5--4) integrated intensity map; the contours 
are at 0.75 Jy beam$^{-1}$ km s$^{-1}$. {\bf (d)} CO(5--4) spectrum at the 
center of the source. The contribution from the continuum is not subtracted
from the line flux.}
\label{Fig}
\end{figure*}

\section{Results}

Signal is detected in the central channels of both the CO(2-1) and CO(5-4)
spectral data sets at 10 and 5 sigmas respectively. The averages of the channels with detected emission were used 
to produce the integrated maps shown on Figs. 1a and 1c. The CO(2-1) integrated
flux is 1.4 Jy km s$^{-1}$. There is no indication for any extension
of the emission at this resolution. The peak of the source is offset by about 
$1^{\prime \prime} $ from the position of the optical source as measured by HST
(\cite{dey99}). This offset
is well within the astrometric accuracy of HST but
if real it may be indicative of a spatial extinction.

The corresponding CO(2-1) 
profile (Fig 1b) is roughly gaussian, with a FWHM of almost exactly 400
km s$^{-1}$, and a central frequency that corresponds to a redshift of $z= 
1.439\pm0.001$.
The appearance of a flat-top or even double peak profile around $\sim 150$ km
s$^{-1}$
(see Fig1b) cannot be checked with the present data and only observations
at higher signal-to-noise ratio (those exploiting the full spectral
resolution) can settle its reality.
The line-width observed is fairly large, although not 
atypical for this kind of sources (e.g. SMMJ02399, Frayer et al. 1998) and 
could be due either to an edge-on system or to various separate components.
The redshift deduced from the CO line is 
apparently shifted from that deduced from the H$\alpha$ line ($z = 1.443$), 
but corresponds to the redshift deduced from the [\OII] line at 3727 \AA\ 
($z=1.439$).
\hfill\break
This shift is still within the uncertainties in the optical redshift
but if real it would be different from what is typically found for low
redshift
luminous galaxies (\cite{san96}), where systematic blue-ward offsets of optical
lines from the CO redshift are attributed to outflows with dust obscuration
(see e.g., \cite{gon98}). The question remains, however, completely open since
a recent analysis by McIntosh et al. (1999)
of a sample of quasars shows how high-redshift objects present
H$\beta$ lines with a systematic mean red-ward shift of $\sim$ 500 km/s
with respect to the systemic redshift of the objects (that defined by the
narrow line region). Even though the comparison with quasars may not be fair
since the line emitting regions could be different,
HR10 seems to show similar properties with the CO redshift corresponding to
that of the narrow forbidden lines and coinciding with the
centre of mass of the system, while the H$\alpha$ line is shifted
with respect to that. 

The integrated map corresponding to the CO(5-4) map (Fig. 1c) shows a source
at the same position as the CO(2-1) integrated map. The integrated intensity 
of that source (corrected for the contribution of the continuum) is 1.35 Jy 
km s$^{-1}$. The corresponding CO(5-4) profile (Fig. 1d) is roughly gaussian, 
with a FWHM of 380 km s$^{-1}$ similar to that of the CO(2-1) profile. The 
central frequency corresponds to a redshift of $z = 1.440\pm0.001$ similar 
to that deduced from the CO(2-1) line.
\hfill\break
Although the CO $J=5$ level is $J(J+1) \cdot 2.77 {\rm K} = 83 {\rm K}$ above
the ground state
the integrated flux (in Jy km s$^{-1}$) of CO(5-4) is equal to
that of the CO(2-1) line, $ {\rm (5-4)/(2-1)} \sim 1$.
CO luminosities, in solar units, are $1.5$ and $3.7 \times 10^7$ L$_\odot$ for the
CO(2-1) and CO(5-4) line respectively, while
the total line luminosities
\LCO\ are $4 \times 10^{10}$ and $6 \times 10^{9}$ K km s$^{-1}$ pc$^2$
$h^{-2}_{\rm 50}$,
for the CO(2-1) and CO(5-4) line respectively. When expressed in these
latter units, the ratio between 
the CO(2-1) and CO(5-4) luminosities of the same source 
is proportional to the line intrinsic brightness (Rayleigh-Jeans) temperature 
ratio integrated over the area of the source:
\begin{equation}
\Re = \frac{T_{\rm b}[{\rm CO(5-4)}]}{T_{\rm b}[{\rm CO(2-1)}]} = \frac{L^\prime_{\rm CO} (5-4)}{
L^\prime _{\rm CO} (2-1)}
\cdot \frac {\Omega_{\rm s} (2-1)}{\Omega_{\rm s} (5-4)}
\end{equation}

\noindent
where $\Omega _{\rm s}$ is the source solid angle.
If the spatial extent of the CO(5-4) emission region
is similar to that of the CO(2-1) -- a plausible hypothesis
since both transitions have same line-width and profile -- $ \Re=0.57$ and
corresponds to a value of the excitation temperature of $T_{\rm ex}
\sim 18$ K (see  e.g. \cite{mal88}).
If gas and dust are in thermodynamic equilibrium the kinetic
temperature $T_{\rm kin}$ would equal $T_{\rm dust}$, but it is usually found that
$T_{\rm kin} < T_{\rm dust}$. In HR10 the dust temperature
was estimated to be $\sim 40 {\rm K}$ (\cite{cim98,dey99}) and this value can be
taken as the upper limit to the gas temperature. If we assume that
$T_{\rm kin} \simeq 20$ K the gas density implied by this ratio is less than
$10^3 {\rm cm} ^{-3}$. As an example, at $T_{\rm rad} = 2.77 (1+z) = 6.76$ K and
$T_{\rm kin} = 20$ K the estimated excitation temperature
is $T_{\rm ex} = 11$ K for a CO density of $ 300 {\rm cm}^{-3}$.
The present data cannot distinguish between a picture
where the dominant component of the ISM in this system is a diffuse
($n(H_{\rm 2}) \sim 10^2 - 10^3 {\rm cm}^{-3}$) gas
or whether the medium is clumpy.
The spatial shift between the CO and H$\alpha$ lines would be more compatible
with this latter picture.

If the ratio between \LCO\ and the mass of molecular gas is
similar in HR10 and Arp 220 (\cite{sco97b}), the 
molecular gas mass in HR10, using the CO(2-1) line, is $ {\rm M (H_{2})} = 
 1.6 \times 10^{11} ~{\rm h} ^{-2}_{\rm 50} {\rm M}_\odot$, larger than what is usually 
found in local ULIRGs (\cite{sol97,bra98}) but 
similar to that of other detected high-z sources (\cite{fra98}).

The 1.3mm continuum map shows only a marginal ($2\sigma$) detection at the 
position of the source, with an integrated flux of $2.2 \pm 0.9$ mJy 
beam$^{-1}$. The flux detected with the IRAM 30m telescope was $4.9\pm
0.8$ mJy (\cite{cim98}) at 240 GHz. Scaling the flux as 
$\nu^{4}$, the PdBI detection would correspond to an expected flux of
$2.4 \pm 1.0$ mJy at the observed frequency of the 30m. We exclude that
this discrepancy is due to an extended component;
it is more likely due to the difference in the calibration of the
two instruments since the two values are consistent within the error bars.

\section{Implications}

The analysis of the spectral energy distribution of HR10 shows
the presence of thermal emission at rest-frame $\lambda > 60 \mu m$
with a range of dust temperatures between 30 and 45 K.
The implied total dust mass is $8-4 \times 10^8 h_{\rm 50}^{-2}$
M$_\odot$ (for a dust emissivity index $\beta$ of 2, \cite{cim98}).
Therefore the resulting gas-to-dust mass ratio for HR10
ranges between 200 and 400, as local spirals (\cite{and95}), ULIRGs (\cite{sol97})
and also sub-mm selected luminous sources show (\cite{fra99}).
\hfill\break
The total rest-frame far-IR luminosity in the range
$10-2000 \mu m$ is $2-2.5 \times 10^{12} h_{\rm 50}^{-2}$ L$_\odot$
(\cite{cim99}) as estimated taking into account the ISO upper limits
at 90 and 170 $\mu m$
(\cite{ivi97}). When these latter are not considered and
the 450 $\mu m$ detection is included the luminosity turns out to be
a factor of 3 larger (\cite{dey99}). 
The ratio $\frac{L_{\rm FIR}}{L^\prime_{\rm CO}}$ lies therefore in the range
$\sim 60-175 L_\odot$ (K km s$^{-1}$ pc$^2$)$^{-1}$, which agrees
with the relation found for nearby luminous galaxies (\cite{san96}),
whose emission is mainly powered by star-formation.
Objects whose FIR emission is
dominated by an AGN -- as the hyperluminous Infrared Galaxies --
show
much larger $\frac{L_{\rm FIR}}{L^\prime _{\rm CO}}$ and do not even show up in CO
(e.g., \cite{eva98}).
This indicates that the overall FIR emission by HR10 is dominated by
star formation.
Assuming that most of the FIR luminosity is due to recent OB star formation
activity, the star formation rate 
turns out to be $ {\rm SFR} = \Psi ~ 10^{-10} L_{\rm FIR} \sim  200-500
h_{\rm 50}^{-2}$ M$_\odot$/yr.\hfill\break
\noindent
Star formation efficiency is usually measured by the ratios
$\frac{L_{\rm FIR}}{M_{\rm H_2}}$ and $\frac{L_{\rm H\alpha}}{M_{\rm H_2}}$ (see e.g.
\cite{you99}). While the former shows indeed quite a high value ($16-44$)
similar to that of merging local systems (\cite{you99}), the latter is of
only 0.007 and very likely indicates a large extinction affecting the H$\alpha$
emission.

With the values above for molecular mass and SFR this active phase of gas
depletion lifetime should have lasted at least:

\begin{equation}
t_{\rm gas} = 4 \times 10^{10} (\frac{L_{\rm FIR}}{L^\prime _{\rm CO}})^{-1} =
(0.2\div0.6) ~10^9 ~~ {\rm yr}
\end{equation}

The large value of gas conversion into stars (with respect to local galaxies)
could be consistent with two possible scenarios:
either a genuinely young galaxy in the process of active star-formation
(and the detected amount of gas seems enough to feed it), or the
presence of a large amount
of gas could be the result of a merging process of two discs
(in this latter case the resulting galaxy will have a mass of
a present-day massive elliptical).

Most of the properties of HR10 suggest that the `locus', which best
characterizes it, is that of local ULIRGs (\cite{hug97}). 
HR10 follows also the expected tight correlation between the infrared flux and
the radio continuum: in fact the
logarithmic ratio of FIR (60 $\mu m$) and radio (1.5GHz) continuum flux density (in 
HR10 rest-frame)
$ q = \log \frac{f_{\rm FIR}}{f_{\rm radio}} \simeq 3$ again falls within the
value of nearby starbursts (\cite{san96}).
Furthermore, the ratio between the line (2-1) and (5-4) 
and the FIR luminosities, 
$\frac{L_{\rm CO(2-1)}}{L_{\rm FIR}} = (2-6) \times  10^{-6} $ and
$\frac{L_{\rm CO(5-4)}}{L_{\rm FIR}} = (0.5-1.5) \times 10^{-5}$, agree with a model
of CO emission in high redshift galaxies, based on an extrapolation of
the properties of local ULIRGs (\cite{bla99}).

With the detected line width and the upper limit on the
CO source size, given by the effective beam width $\theta < 5
^{\prime \prime}$,
the upper limit to the total dynamical mass contained within the CO
emitting region is:

\begin{equation}
M_{\rm dyn} = \frac{R}{2 ~ G} ~ (\frac{\Delta V}{\sin i})^2 < 
7.7 \times  10^{11} (\sin i)^{-2} h_{\rm 50}^{-1}M_\odot
\end{equation}

\noindent
where $\Delta V$ is the observed deconvolved line width, $i$ is the
inclination
and $R$ is the linear diameter of the source ($R < 43 h_{\rm 50}^{-1} kpc$).
The resulting dynamical mass is a factor of 5 larger than
the estimation of the molecular mass. The two values would coincide
if the CO emission were concentrated within the inner
10 kpc.

\begin{acknowledgements}
We are grateful to Roberto Neri for his help during
the data reduction and analysis,
all the IRAM staff for providing us with excellent data and the referee,
Francoise Combes, for helping in the interpretation of these observations.\hfill\break
Part of this work was supported by the CNR contract {\it progetto strategico
"Astronomia Submillimetrica" del Comitato Scienze Fisiche}.
P.A. thanks MPE for hospitality during 1999 when this work was done.\hfill\break
We dedicate this paper to the memory of the 25 people who lost their lives
in the tragedies of the Plateau de Bure cable-car and helicopter.
\end{acknowledgements}

\end{document}